\documentclass[a4paper,11pt]{article}

\usepackage[T1]{fontenc} 

\usepackage[english]{babel}
\usepackage{amsmath,amssymb}
\usepackage{jcappub} 
\usepackage{graphicx}

\def\nat{Nature}

\def\prd{Phys. Rev. D}
\def\prl{Phys. Rev. Letters}
\def\mnras{MNRAS}
\def\apj{ApJ}
\def\apjl{ApJL}
\def\apjs{ApJS}
\def\aap{A\&A}

\def\araa{ARA\&A}

\def\physrep{Phys. Rep.}
\def\pasa{Publ. Astr. Soc. Australia}

\def\jcap{JCAP}

\usepackage[english]{babel}
\usepackage{amsmath,amssymb}

\usepackage{graphicx}

\usepackage[normalem]{ulem}

\usepackage{verbatim}
\usepackage{color}
\usepackage{multirow}
\usepackage{mathtools}
\usepackage{epstopdf}

\usepackage{hyperref}
\usepackage{url}
\usepackage{xcolor}

\usepackage{gensymb}

\usepackage[normalem]{ulem}

\usepackage{orcidlink}
\usepackage{verbatim}
\usepackage{color}
\usepackage{multirow}
\usepackage{mathtools}
\usepackage{epstopdf}

\usepackage{hyperref}
\usepackage{url}
\usepackage{xcolor}

\usepackage{gensymb}

\usepackage{appendix}
\usepackage{verbatim}

\title{Primordial black holes as supermassive black hole seeds}

\author[1,a]{F. Ziparo\orcidlink{0000-0001-6316-1707}\note{Corresponding author.}}
\author[a]{, S. Gallerani\orcidlink{0000-0002-7200-8293}}
\author[a]{, A. Ferrara\orcidlink{0000-0002-9400-7312}}

\affiliation[a]{Scuola Normale Superiore, Piazza dei Cavalieri 7, I-56126 Pisa, Italy}

\emailAdd{francesco.ziparo@sns.it}

\abstract
{
The presence of supermassive black holes (SMBHs, $M_{\bullet}\sim 10^{6-10}~M_{\odot}$) in the first cosmic Gyr ($z\gtrsim 6$)  challenges current models of BH formation and evolution. We propose a novel mechanism for the formation of early SMBH seeds based on primordial black holes (PBHs).  We assume a non-Gaussian primordial power spectrum as expected in inflationary models; these scenarios predict that PBHs are initially clustered and preferentially formed in the high-$\sigma$ fluctuations of the large-scale density field, out of which dark matter (DM) halos are originated. 
Our model accounts for (i) PBH accretion and feedback, (ii) DM halo growth, and (iii) gas dynamical friction. PBHs lose angular momentum due to gas dynamical friction, sink into a dense core, where BH binaries form and undergo a runaway merger, eventually leading to the formation of a single, massive seed. This mechanism starts at $z\sim 20-40$ in rare halos ($M_h\sim 10^7\ M_\odot$ corresponding to $\sim 5-7\sigma$ fluctuations), and provides massive ($\sim 10^{4-5}~ M_{\odot}$) seeds by $z\sim 10-30$. We derive a physically-motivated seeding prescription that provides the mass of the seed, $ M_{\rm seed}(z)=3.1\times 10^{5}\ { M_{\odot}}[(1+z)/10]^{-1.2}$, and seeded halo, $ M_{h}(z)=2\times 10^{9}\ {M_{\odot}}[(1+z)/10]^{-2}e^{-0.05z}$ as a function of redshift. 
This seeding mechanism requires that only a small fraction of DM is constituted by PBHs, namely $f_{\rm PBH}\sim 3 \times 10^{-6}$.
We find that $z\sim 6-7$ quasars can be explained with $6\times 10^4 M_{\rm \odot}$ seeds planted at $z\sim 32$, and growing at sub-Eddington rates, $\langle\lambda_{\rm E}\rangle\sim 0.55$. The same scenario reproduces the BH mass of GNz11 at $z=10.6$, while UHZ1 ($z=10.1$) and GHZ9 ($z=10$) data favour instead slightly later ($z\sim 20-25$), more massive ($10^5~M_{\rm \odot}$), and efficiently accreting ($\langle\lambda_{\rm E}\rangle\simeq0.9$) seeds.  
During the runaway phase of the proposed seed formation process, PBH-PBH mergers are expected to
copiously emit gravitational waves. These predictions can be tested through future Einstein Telescope observations and used to constrain inflationary models.
}


\begin{document}
\maketitle
\flushbottom

\section{Introduction}
Observations in the local and nearby Universe show that supermassive black holes (SMBHs, $M_{\bullet}\sim 10^{6-10}~M_{\odot}$) reside in the center of most galaxies, including the Milky Way \citep[][and references therein]{kormendy2013}. Furthermore, observations of $z\sim 6-7.5$ quasars imply that SMBHs already exist in the first cosmic Gyr of the Universe \citep[][and references therein]{fan2023}. Finally, recent James Web Space Telescope (JWST) data \citep[e.g.][]{GNZ11, UHZ1, Bogdan23, GHZ9} reveal the presence of SMBHs at even higher redshifts ($ z\geq 10$). The existence of such early SMBHs raises thorny questions about the formation and the subsequent growth of the seeds from which these extreme compact objects have been originated \citep[e.g.][]{Volonteri03, Volonteri12, latif2016, Gallerani17,valiante2017,volonteri2021}.

Several scenarios have been proposed for the formation of SMBH seeds: {\it light seeds} ($M_{\rm seed}\sim 10-100\ M_{\odot}$), remnants of massive, metal-free (Population III) stars, originated in early ($z\sim 30$) mini-halos \citep[$M_h\sim 10^6\ M_{\odot}$, e.g.][]{Madau98, Heger03, Yoshida2006simo}; {\it intermediate seeds} ($M_{\rm seed}\sim 10^{2-3}\ M_{\odot}$), produced by nuclear star clusters (NSC) in high redshift ($z\sim 15-20$) galaxies \citep[e.g][]{Devecchi09, Davies11, Lupi14}; {\it heavy seeds} ($M_{\rm seed}\sim 10^{4-6}\ M_{\odot}$), also called direct collapse black holes (DCBHs), formed through the collapse of atomic, pristine gas in high redshift ($z\sim 8- 17$) atomic cooling ($M_h\gtrsim10^8\ M_{\odot}$) halos \citep[e.g.][]{Eisenstein95, Silk98, Ferrara14}. Whether or not these scenarios can explain the SMBH masses found in $z\gtrsim 6$ quasars and recent JWST data depends on the accretion history that follows their birth, as parametrized by the Eddington ratio $\lambda_E=\dot{M}/\dot{M}_E$, where ($\dot{M}_E$) $\dot{M}$ is the (Eddington-limited) accretion rate. 
 
The {\it light seeds} scenario certainly provides the most natural path for the formation of SMBH seeds. However, to explain the existence of SMBHs at high-$z$, PopIII remnants require sustained ($\lambda_{E}>1$) accretion rates for prolonged ($\sim \rm Gyr$) intervals of time. On the contrary, the {\it intermediate} and {\it heavy seeds} scenarios have the advantage that the produced seeds can grow at sub-Eddington rates ($\lambda_{E}<1$). However, the conditions for the formation of seeds from NSCs and DCBHs are not easily satisfied. Given the criticalities associated to each of the aforementioned scenarios \citep{Volonteri03, Lodato06, Tanaka09, Volonteri12}, the quest for candidates of SMBH seeds has also involved exotic particles, such as primordial black holes (PBHs).

PBHs are expected to form in the radiation dominated era ($t_{\rm H}\sim 1$ s,  \citep{Carr74}) due to the gravitational collapse of overdense regions, and have been proposed as non-baryonic dark matter (DM) candidates \citep{Chapline75}. The fraction of dark matter composed by PBHs ($f_{\rm PBH}$) have been constrained by exploring their impact on several astrophysical phenomena (see \citep{Carr20}), such as gamma rays emission from evaporating PBHs \citep{Laha19,Coogan21}, microlensing effects \citep{Niikura19,Blaineau22} and disruption of wide binaries or ultra-faint dwarfs \citep{Monroy14,Brandt16}. Furthermore, signatures of accreting PBHs are expected to be left on the CMB spectrum and its anisotropies \citep{Poulin17}, the 21 cm power spectrum \citep{Mena19}, and the X-ray/radio/infrared  backgrounds \citep{Cappelluti22, Zip22, Manzoni23}. Although current constraints on $f_{\rm PBH}$ are quite stringent \citep[typically $f_{\rm PBH}\lesssim10^{-2}-10^{-3}$ for $10^{-10}<M_{\rm PBH}<10^{15}~\rm M_{\odot}$, see Fig. 10 by][]{carr2021}, this does not question that PBHs could have formed in the early Universe and affect the subsequent evolution of structure formation \citep[e.g.][]{Inman2019}. 

In particular, it has been proposed that PBHs may constitute the seeds of SMBHs \citep[e.g.][]{Duchting04, Kawasaki12, Dayal2024}. PBHs formed with low masses ($ M_{\rm PBH}\sim 100\ M_{\odot}$), can mimic the {\it light seeds} scenario; in this case, however, they would have more time to accrete with respect to PopIII remnants \citep{Bernal18}. Alternatively, if originated during inflation, they might be as massive as the seeds expected in the \textit{intermediate/heavy} seeds scenarios \citep[$ M_{\rm PBH}\sim 10^3-10^5\ M_{\odot}$, e.g.][]{Hasinger20, Cappelluti22}. Furthermore, PBHs may participate to the growth of stellar compact remnants (neutron stars and stellar-mass black holes) while migrating towards the galactic nuclear region, because of gas dynamical friction \citep{Boco20, Boco21}.

In this work, we propose a new seeding mechanism based on 30~$\rm M_{\odot}$ PBHs, distributed in the very central region ($ \lesssim 1\ \rm pc$) of high-$z$ ($z>20$) rare ($\sigma\geq 5$) DM halos. The paper is organized as follows: in Sec. \ref{Sec:02}, we describe the semi-analytic model we developed to produce SMBH seeds; in Sec. \ref{Sec:03}, we present the outcomes of our model, in terms of seed masses and epochs of their formation, and we explore the implications of our seeding mechanism on the formation of early SMBHs; in Sec. \ref{Sec:04}, we summarize the main ingredients and findings of our model and we discuss caveats of our results.

\begin{figure*}
\includegraphics[width=1\textwidth]{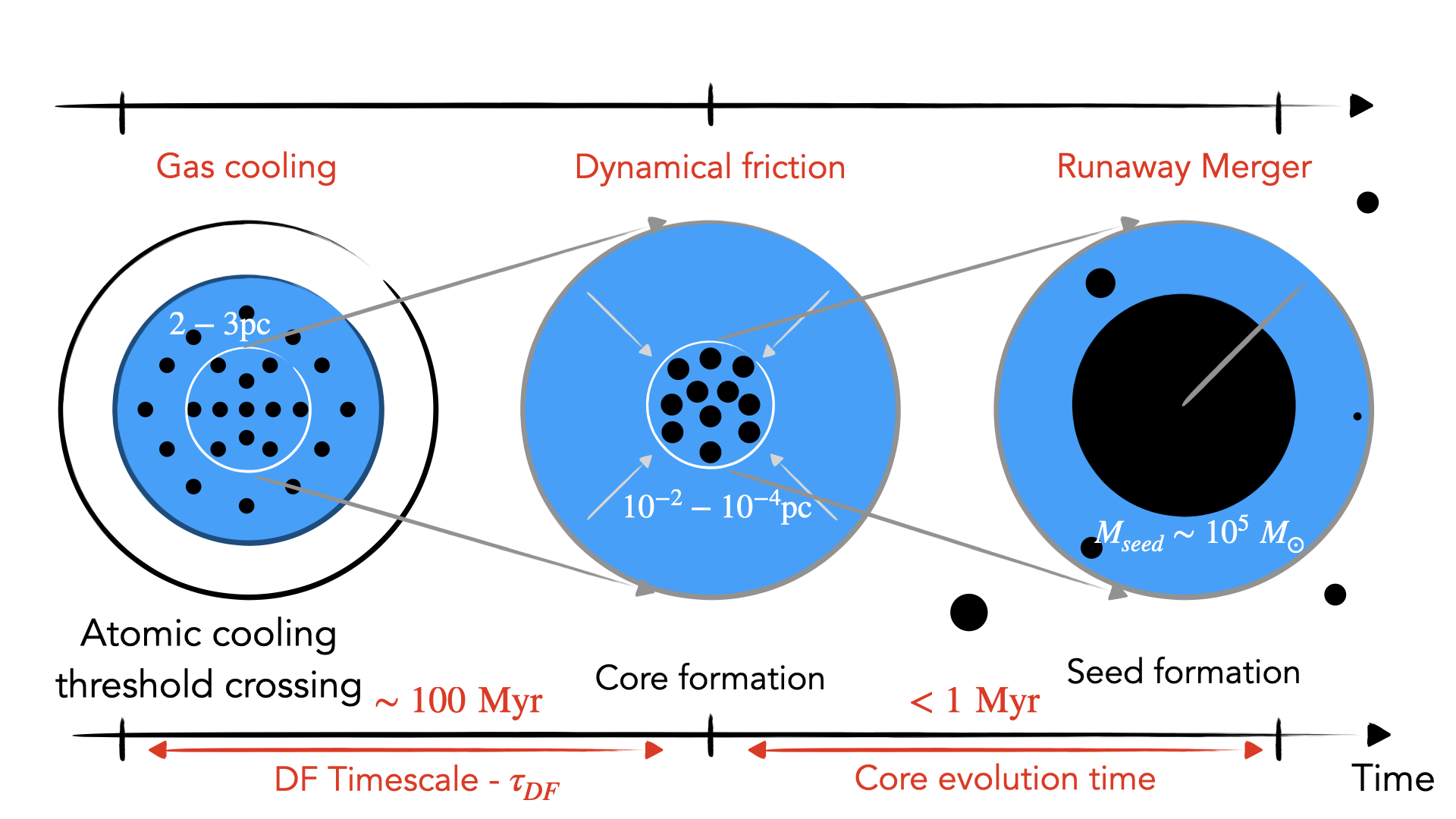}
\caption{Schematic description of the seed formation process presented in this work. High density peak in the early Universe collapse forming dark matter halos comprised of PBHs. The halo mass increases as cosmic time progresses, reaching a point where its temperature surpasses the atomic cooling threshold. This initiates the gas cooling process within the halo. As the gas cools, it becomes denser and settles towards the center of the dark matter gravitational potential, embedding the PBHs at the center. PBHs accreting and moving into dense gas lose angular momentum due to dynamical friction and gather in a core. At the moment that the core is dense enough to retain the products of binary merging, it starts to evolve. PBH binaries form and merge extremely rapidly due to the high density, ultimately resulting in a runaway merger phase. The core is finally converted into a seed.}\label{sketch}
\end{figure*}
\section{Model}\label{Sec:02}
In this section, we describe the model we use to investigate whether PBHs can be considered as valid candidates for SMBH seeds.  Fig. \ref{sketch} shows a schematic view of the model: we start from a configuration in which PBHs are initially clustered (Sec. \ref{Sec:02.0}) and represent a fraction $ f_{\rm PBH}=\Omega_{\rm PBH}/\Omega_{\rm dm}$ of dark matter that coexists with baryons (Sec. \ref{Sec:02.1}); in Sec. \ref{Sec:02.3} we describe how DM halos grow and how baryons settle in their potential wells in the form of cooled gas;
PBHs both accrete baryons (Sec. \ref{Sec:02.2}) and loose angular momentum as a consequence of the dynamical friction on the gas (Sec. \ref{dynamicalfriction}), thus gathering in the central region of the potential well and forming a dense core; 
PBHs clustered in the core undergo a runaway collapse that ends up into a massive BH whose mass depends on the initial conditions of the configuration (Sec. \ref{Sec:02.5}).       
\subsection{PBH clustering from primordial non Gaussianities} \label{Sec:02.0}
Let us consider a simplified picture of the Universe in which primordial perturbations consist of only two components: short- and long- wavelength modes, out of which PBHs and large-scale structures, respectively, are formed \citep{sasaki2018}. Let us assume that the primordial power spectrum is characterised by local non-Gaussianities that are small enough not to violate CMB constraints ($f_{\rm NL}^{\rm local}\geq-0.9\pm 5.1$, \citep{PlankNG}). A non-vanishing $f_{\rm NL}^{\rm local}$ is predicted by inflationary models \citep[e.g.][]{Bartolo2004} and leads to a coupling between long- and short-wavelength modes \citep{Young_Byrnes2}. In this case, the amplitudes of the short modes are coupled/modulated by the long modes, being larger (smaller) at the peak (in the trough) of the long-wavelength mode \citep[see Fig. 1 in][]{Tada15}. In this framework, PBHs are initially clustered and preferentially formed in the high-$\sigma$ fluctuations of the large-scale density field, out of which DM halos are originated \citep{Young15}. 
\subsection{PBH, DM and gas distribution}\label{Sec:02.1}
Given the discussion presented in \ref{Sec:02.0}, we assume that PBHs represent the {\it whole} of DM, but {\it only} in collapsed haloes. In other words, $f_{\rm PBH}=1$ for $r<r_{\rm vir}$, where $r$ is the distance from the center of a DM halo and $r_{\rm vir}$ is its virial radius, and $f_{\rm PBH}=0$ elsewhere\footnote{We further discuss this assumption in Sec. \ref{fPBH}.}. The number of PBHs as a function of $r$ can be then written as:
\begin{equation}
N_{\rm PBH}(r)=\frac{4\pi}{M_{\rm PBH}}\int^{r}_{\rm 0}\rho_{\rm dm}(r')r'^2dr',
\end{equation}
where\footnote{We assume a
monochromatic PBH mass function and we defer to a future work the investigation of extended mass functions.} $ M_{\rm PBH}=30\ M_{\odot}$ is the PBH mass \citep[e.g.][]{Zip22}, and $\rho_{\rm dm}$ is the standard Navarro, Frenk \& White \citep[NFW,][]{NFW97} density profile: 

\begin{equation}
\rho_{\rm dm}(r)=\frac{\rho_c\delta_c}{cx(1+cx)^2},
\end{equation}
where $ \rho_c$ is the critical density of the universe, $ \delta_c$ is the critical density for the collapse, $ x=r/r_{\rm vir}$ denotes the radial distance in units of the virial radius $ r_{\rm vir}$.

The virial radius and temperature of a DM halo of mass $ M_{h}$ can be computed as \citep{Barkana01}:
\begin{equation} 
\begin{split}
  r_{\rm vir}=\ &0.784\ \left(\frac{M_{h}}{10^8\ h^{-1}\ M_{\rm \odot}}\right)^{1/3}\left[\frac{\Omega_m}{\Omega^z_m}\frac{\Delta_c}{18\pi^2}\right]^{-1/3}\ \\ 
  &\left(\frac{1+z}{10}\right)^{-1}h^{-1}\ \mathrm{kpc},
\end{split}
\end{equation}
\begin{equation} 
\begin{split}\label{Tvir}
T_{\rm vir}=\ & 1.98\times 10^4\ \left(\frac{\mu}{0.6}\right)\ \left(\frac{M_{\rm vir}}{10^8\ h^{-1}\ M_{\rm \odot}}\right)^{2/3}\ \\ & \left[\frac{\Omega_m}{\Omega^z_m}\frac{\Delta_c}{18\pi^2}\right]^{1/3}\left(\frac{1+z}{10}\right)\ \rm K,
\end{split}
\end{equation}
 where the overdensity\footnote{We adopt a $ \Lambda$CDM cosmology in agreement with Planck18 \citep{Planck18} results: $ \Omega_{m}=0.315$, 
$\Omega_{\rm \Lambda}=0.685$, $\Omega_{b}=0.049$, $\sigma_8 = 0.811$, $n_s=0.965$, 
and $H_{\rm 0}=100 \,h$~km~s$^{-1}$~Mpc$^{-1}=67.4$~km~s$^{-1}$~Mpc$^{-1}$.} relative to $\rho_c$ at the collapse redshift can be expressed as $\Delta_c=18\pi^2+82d-39d^2$, with $d=\Omega_m^z-1$ and $\Omega_m^z=\Omega_m(1+z)^3/(\Omega_m(1+z)^3+\Omega_{\Lambda})$; $\Delta_c$ is related to $\delta_c$ through the following relation:

\begin{equation}\label{overD}
\delta_c=\frac{\Delta_c}{3}\frac{c^3}{ln(1+c)-c/(1+c)},
\end{equation}
where the concentration parameter $ c$ is computed following the model described in \citep{Prada12}. 

We assume that the gas is distributed following a singular isothermal radial density profile and is in hydrostatic equilibrium with dark matter \citep{Makino98}:
\begin{equation}\label{makino}
    \rho_g(r)=\rho_0 \exp \biggl({-\frac{\mu m_p}{2k_BT_g}[v^2_e(0)-v^2_e(r)]} \biggl),
\end{equation}
where $ \rho_0$ is obtained normalizing the total mass integrated over the profile to the gas mass $M_g=(\Omega_b/\Omega_m)M_h$, and $T_g=T_{\rm vir} (M_h,z)$. 

\subsection{Halo growth and gas cooling}\label{Sec:02.3}
As time passes, DM halos grow through minor/major mergers and accretion from the IGM, and their mass evolution can be described through the following relation \citep{Fakouri10}:
\begin{equation}\label{Fakouri}
\begin{split}
    \frac{dM_h}{dt}= & 46.1\biggl(\frac{M_h}{10^{12}M_{\odot}}\biggl)^{1.1}(1+1.11z) \\ &  \sqrt{\Omega_m(1+z)^3+\Omega_{\Lambda}} M_{\odot}\rm yr^{-1}.
\end{split}
\end{equation}

When $M_h$ becomes large enough to have a virial temperature above the atomic cooling threshold\footnote{The assumed value $T_g=10^4\rm K$ is motivated by the gas composition (atomic hydrogen) at early epochs ($ z\gtrsim 10$). The hydrogen cooling function has a sharp drop below this value \citep{Katz96}.} $T_g=10^4\rm K$, a fraction $f_g=0.1$ of the gas mass cools down to the same temperature\footnote{We expect the gas to be fully ionized by PBHs feedback in the central zone, and we assume a temperature of $10^4\rm \ K$ for the ionized gas \citep{Anderson09}.} and reaches high density values ($n_{g,0}\sim 10^2-10^5\ \rm cm^{-3}$), that favors the BH accretion and starts the dynamical friction process described below. Hereafter, we call "crossing redshift" ($z_{\times}$) and "crossing mass" ($M_{h,\times}$) the redshift and the halo mass when $T_{\rm vir}(M_h,z)=10^4$~K. We consider these values as the initial conditions for the calculations described below.

\subsection{PBHs accretion} \label{Sec:02.2}
PBHs accrete the surrounding gas following the Bondi-Hoyle-Lyttleton prescription \citep{hoylelyttleton, BondiHoyle}:
\begin{equation}\label{bondi}
    \dot{M}_{\rm PBH}=\frac{G^2M^2_{\rm PBH}\rho_g}{(c_s^2+v_{\rm PBH}^2)^{3/2}},
\end{equation}
where $ c_s=( k_BT_{\rm g, c}/\mu m_p)^{1/2} {\rm km\ s^{-1}}=8.3\ \rm km\ \rm {s}^{-1}$ is the sound speed,
and $ v_{\rm PBH}$ is the relative velocity between gas and PBHs, here approximated to the circular velocity at a radius $r$ as: 
\begin{equation}
    v_c(r)=\sqrt{\frac{G(M_g(r)+M_{\rm dm}(r))}{r}}.
\end{equation} 
The accretion rate is limited at the Eddington rate:
\begin{equation}
    \dot{M}_{\rm E}=\frac{L_{\rm E}}{\varepsilon c^2},
\end{equation}
where $ L_{\rm E}=1.26\times (M_{\rm PBH}/M_{\odot})L_{\odot}$ is the Eddington luminosity, $ \varepsilon=0.1$ is the radiative efficiency, and $ c$ is the speed of light. 
The radiation emitted in this process provide thermal and ionization feedback \citep{Hasinger20} strong enough to prevent $ H_2$ formation, and therefore to hamper star formation. Since the volume filling factor of PBHs Str{\"o}mgren spheres is O(1) in the central region of the halo ($ r\sim \rm pc$), we consider the gas to be fully ionized ($T_g=10^4$K).

\begin{figure}
\begin{center}
\includegraphics[width=.8\textwidth]{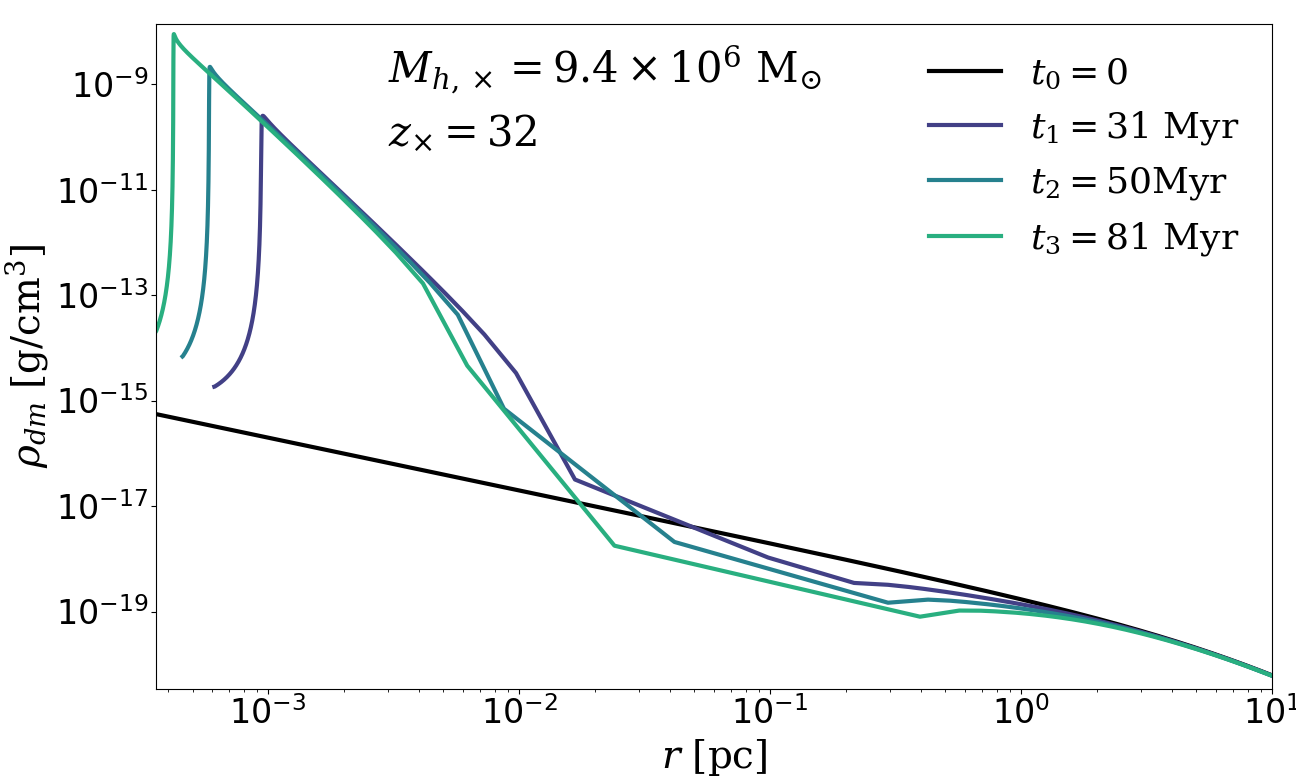}
\caption{Dark matter density profile evolution with time. The black line ($ t_0=0$) shows a classic NFW profile \citep{NFW97}. The purple, blue and green lines show the profile evolution at time $ t_1=31\ \rm{Myr}$, $ t_2=50\ \rm{Myr}$, $ t_3=81\ \rm{Myr}$. The plot shows a DM density spike in the innermost part of the halo. This spike is caused by PBHs gradually shrinking their orbits due to gas dynamical friction.}
\label{nfw_evo}
\end{center}
\end{figure}

\subsection{Dynamical Friction}\label{dynamicalfriction}
We treat the dynamics of a system of black holes orbiting in the halo potential and embedded in the halo gas, following  \citep{Mo2010}. Angular momentum loss of PBHs on the gas\footnote{In contrast to what happens when a black hole moves through neutral gas, in fully ionized medium \citep{Park17} and for a black hole in supersonic motion \citep{Ostriker99} the dynamical friction can work even in the absence of stars.} reduces the orbits, bringing the black holes to form a dense cluster. 
The equation that regulates this process is:
\begin{equation}
    \frac{dL}{dt}=-0.428 \ln \Lambda\frac{GM_{\rm PBH}}{r},
\end{equation}
where $ L$ is the specific angular momentum, and $ \Lambda$ is the Coulomb logarithm that depends on $ M_{\rm PBH}$ and $ M_{g}$ through the following expression: 
\begin{equation}
\label{coulumb_log}
    \ln \Lambda = \frac{1}{2}\ln\biggl[1+ \biggl(\frac{b_{\rm max}}{b_{90}}\biggl)^2\biggl],
\end{equation}
where $ b_{\rm max}$ is the maximum impact parameter, and $ b_{90}$ is the impact parameter for which the deflection angle of the reduced particle of the encounter is equal to $90^\circ$.
Following \citep{BinnTrem08}, we define $ b_{\rm max}=r$, where $ r$ is the radius of the circular orbit assumed for PBHs, and $ b_{90}=v^2_{\rm c}/GM_{\bullet}$. 
Eq. \ref{coulumb_log} can thus be rewritten as:
\begin{equation}
    \ln \Lambda = \frac{1}{2}\ln\biggl[1+ \biggl(\frac{M_g+M_{\rm dm}}{M_{\rm PBH}}\biggl)^2\biggl].
\end{equation}

The dynamical friction on the gas has the effect of shrinking the radius of the PBH orbit with a rate $ dr/dt$ such that:
\begin{equation}\label{r_decay}
    r\frac{dr}{dt}=-0.428\ln\Lambda\frac{GM_{\rm PBH}}{v_c}.
\end{equation}
This expression can be used to compute the dynamical friction timescale,
\begin{equation}
    \tau_{\rm df}=\frac{1.17}{\ln \Lambda}\frac{r^2_i v_c}{GM_{\rm PBH}}.
\end{equation}
\subsection{Runaway merger}\label{Sec:02.5}
Due to gas dynamical friction PBHs migrate towards the center of the gas ($ r\leq 0.01\rm \ pc$) forming a cluster of PBHs. 
As a consequence of the PBH clustering, the distance between PBHs can become small enough for PBHs to form binaries and undergo runaway mergers, possibly leading to runaway merger that ends up with the formation of a single, massive black hole, which would represent the seed for a SMBH.

\begin{figure}
\centering
\includegraphics[width=1\textwidth]{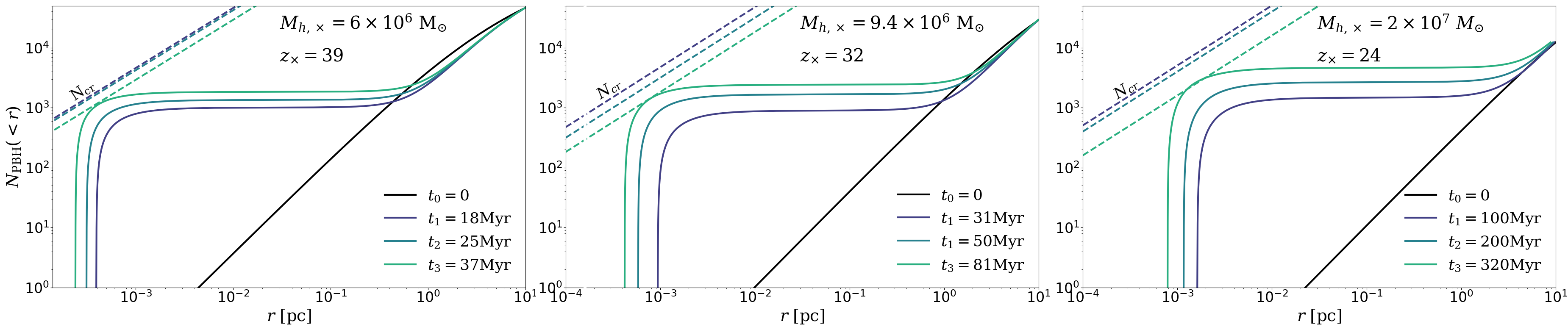}
\caption{Time evolution of the PBH cumulative number within 10 pc of the DM halo center. Different panels represent different cases of the crossing redshift $z_{\times}$ and halo mass $M_{h,\times}$, as labelled in the figure. The black straight line represents the initial PBH distribution following the NFW profile; curved coloured lines show how the PBH distribution evolves with time due to gas dynamical friction; dashed straight lines denote the critical PBH cumulative number $ N_{\rm cr}$ required to reach the escape velocity $ v_{\rm esc}=1000\ \rm kms^{-1}$. 
}\label{triplevo}
\end{figure}

For this process to occur it is necessary that, after mergers, the products are retained within the core, namely that PBHs resulting from mergers are characterised by a recoil velocity $v_k$ smaller than the escape velocity $ v_{\rm esc}$ of the core \citep{Davies11}. The kick velocity that a PBH can attain after the merging with another PBH, as a consequence of the asymmetric emission of gravitational radiation, depends on the mass ratio of the merging PBHs and on their spins. We adopt as a fiducial maximum value $ v_k^{\rm max}=1000\ \rm km s^{-1}$ \citep{Davies11} and, in Appendix \ref{Rapster}, we quantify how much this assumption affects our results. Once that $v_k$ is fixed, the condition for the runaway merger is given by:
\begin{equation}\label{critical}
    v_{\rm esc}(r_{c})=\sqrt{\frac{GM_{c}(r_{c})}{r_{c}}}\geq v_{k},
\end{equation}
where $r_c$ and $M_c$ represent the radius and the mass of the core, respectively. 

These two quantities evolve with time, as a consequence of the PBH clustering and accretion processes that cause $r_c$ to shrink and $M_c$ to increase. This effect can be visualised in Fig. \ref{nfw_evo} that shows how the dark matter density profile evolves with time. Initially, the PBH distribution follows the NFW profile (solid black line). As time passes, PBHs accrete and sink in the central region, forming a core that becomes increasingly smaller and more massive (solid coloured lines), thus being characterised by an increasing escape velocity. In particular, $v_{\rm esc}$ can reach at a given time the critical value of $v_k^{\rm max}$, thus satisfying the condition for the runawy merger into a single BH. The mass of the BH resulting from this process is given by $M_{\rm seed}=N_{\rm PBH}(r_c)\times M_{\rm PBH}$, where $N_{\rm PBH}$ is the number of PBHs enclosed in $r_c$, determined as follows.

We first compute how the number of PBHs as a function of the radius $N_{\rm PBH}(r)$ evolves with time (solid curved lines in Fig. \ref{triplevo}, for different crossing halos and redshifts). We find that in $\sim 10-100$~Myr, the dynamical friction boosts the PBH number in the core, shrinking PBHs orbits from $<10$~pc down to $\sim 10^{-3}$~pc. We then define $N_{\rm cr}$ as the {\it critical number} of PBHs such that 
\begin{equation}\label{critical}
    v_{esc}(r)=\sqrt{\frac{G N_{\rm cr}(r)\times M_{\rm PBH}}{r}}= v_{k}^{\rm max},
\end{equation}
shown with dashed lines in Fig. \ref{triplevo}. We underline that $N_{cr}$ increases linearly with the radius and decreases with time. The time dependence is due to the fact that PBHs get more massive because of the accretion process; therefore, a smaller number of them is enough to verify the condition in Eq. \ref{critical}. Fig. \ref{triplevo} shows that, at a given time, it can be identified the core radius $r_c$ as the distance from the center that satisfies the following condition: $N_{\rm PBH}(r_c)\geq N_{\rm cr}(r_c)$.

In halos with crossing masses and redshifts such that the condition in Eq. \ref{critical} is verified during their evolution, the seed mass resulting from the runaway merger of PBHs in the core is given by $M_{\rm seed}=N_{\rm PBH}(r_c)\times M_{\rm PBH}=M_{c}(r_{c})$.

\begin{figure*}
\includegraphics[width=1\textwidth]{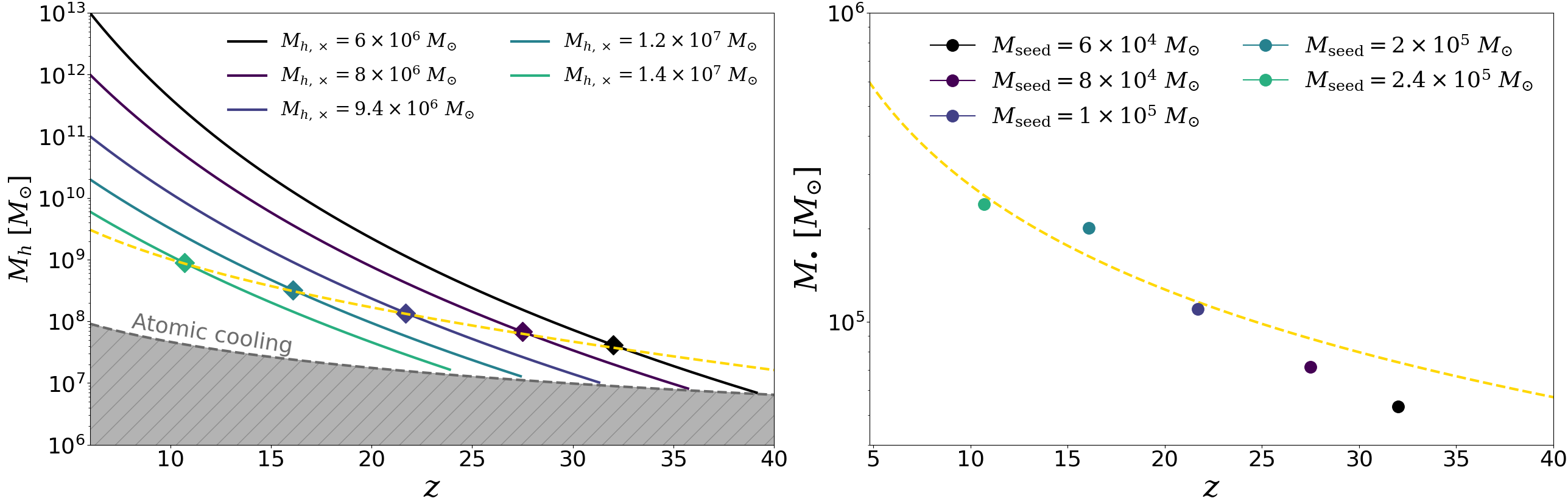}
\caption{\textit{Left panel:} Redshift evolution of DM halos characterized by $M_h\sim 10^{10-13}~M_{\odot}$ masses at $z=6$ and crossing masses $M_{h,\times}\sim 10^{6-7}~M_{\odot}$ at $z_{\times}=25-40$. Different lines show different $M_{h,\times}$ and $z_{\times}$ combinations, as labelled in the figure. Diamonds represent the DM halo mass at the time when the runaway merger occurs and the massive seed is formed. The golden dashed line denotes the best fit for the redshift evolution of the halo masses at the seed formation epoch (Eq. \ref{host_mass}). The shaded grey area represents the $z-M_h$ parameter space below the atomic cooling threshold. Perturbations that are above the atomic cooling threshold at earlier times are larger and produce less massive seeds in shorter time, as detailed in the main text. \textit{Right panel:} Seed mass evolution with redshift. Colored filled circles show the mass of different seeds associated to the halo masses (diamonds) in the right panel. The golden dashed line represents the best fit for the redshift evolution of the seed masses (Eq. \ref{seed_mass}). }\label{Seedshalo_evo}
\end{figure*}

\section{Results}\label{Sec:03}
In this section, we present the results obtained by solving the three coupled differential equations regulating the PBH accretion (Eq. \ref{bondi}), the evolution of the halo mass (Eq. \ref{Fakouri}), and the shrinking of the PBH orbits (Eq. \ref{r_decay}). We first quantify the seed masses, and the masses and redshifts of the host DM halos (Sec. \ref{seeds_form}); then, we provide instructions about the seeding prescription resulting from our work (Sec. \ref{seed_prescription}), which can be used as an input of semi-analytical models and cosmological simulations; furthermore, we quantify the fraction of DM into PBHs ($f_{\rm PBH}$) required for this seeding mechanism to work (Sec. \ref{fPBH}); finally, we check whether our model can provide a viable seeding mechanism to explain observations of $6<z<11$ SMBHs (Sec. \ref{implications}). 
\subsection{Seed formation}\label{seeds_form}
Fig. \ref{Seedshalo_evo} shows the redshift evolution of DM halos that reach masses $\sim 10^{10-13}~M_{\odot}$ at $z=6$, corresponding to $5-7\sigma$ fluctuations of the density field. We focus on this specific redshift range because the main goal of our work is to explain the existence of SMBHs at $z\gtrsim 6$. The figure shows that the crossing mass does not vary much with the $\sigma$ fluctuation. For what concerns the crossing redshift, we find that the smaller is the $\sigma$ fluctuation the lower is the crossing redshift, namely the redshift when the halo mass becomes larger that the atomic cooling mass threshold (e.g. $z\sim 20-40$ for $\sigma=5-7$, respectively). This is due to the fact that lower $\sigma$ fluctuations are characterised by longer dynamical friction timescales ($\tau_{df}\sim 10-100\ \rm Myr$ for $\sigma=5-7$). 

Such differences arise from different initial conditions in the central part ($r\sim 10\  \rm pc$) of the halos at the crossing epoch. For $\sigma\simeq7$, the high central gas density ($ n_{g, 0}\sim 10^5 \rm{cm^{-3}}$) implies efficient accretion and dynamical friction; furthermore, the high concentration ($ c=31$) corresponds to a larger number of PBHs in the central zone. For $\sigma\simeq5$, the much smaller central density ($ n_{g, 0}\sim 10^3 \rm{cm^{-3}}$) makes the accretion and dynamical friction less efficient; furthermore, the lower concentration ($c=8$) implies that a longer time is required to gather a number of PBHs large enough to reach the critical number $N_{\rm cr}$ (see Eq. \ref{critical}).

We start the computation considering the crossing mass and redshift as initial conditions. At each timestep, we check if and when the condition in Eq. \ref{critical} is satisfied. The dots overplotted on the continuous lines in Fig. \ref{Seedshalo_evo} denote the mass and redshift when this takes place. Since the core collapse occurs on a very short timescale ($ \leq 1~\rm Myr$), we identify the seed formation with the same redshift, and we report for each case the seed mass produced. Our model predicts the formation of massive seed ($M_{\bullet}\sim 10^{4-5}~M_{\rm \odot}$) in high-z ($10<z<30$) DM halos with masses $M_h\sim 5\times 10^7-10^9~M_{\rm \odot}$.

We expect that the seed formation process described in this work naturally stops at $ z\simeq 20$ due to the low gas and PBHs density in the halo center at the crossing time. For example, we verified that perturbations that cross the atomic cooling mass at $z\lesssim22$ do not form a seed even after $ 1\ \rm Gyr$.

\subsection{Seeding prescription}\label{seed_prescription}
We exploit the results reported in Sec. \ref{seeds_form} to obtain a seeding recipe for theoretical models and simulations. 

The diamonds in Fig. \ref{Seedshalo_evo} (right panel) represent the masses that halos have at the time of the seed formation. By fitting these data (yellow dashed), we find that the mass of halos that produce a seed evolves with redshift following this relation:
\begin{equation}\label{host_mass}
    M_{h, \rm seed}(z)=2\times 10^{9}\ M_{\rm \odot}\biggl(\frac{1+z}{10}\biggl)^{-2}e^{-0.05z}.
\end{equation}
Similarly, dots in Fig. \ref{Seedshalo_evo} (left panel) denote how the mass of the seed formed changes with redshift, which can be fitted (yellow dashed) by the following equation:
\begin{equation}\label{seed_mass}
    M_{\rm seed}(z)=3.1\times 10^{5}\ M_{\rm \odot}\biggl(\frac{1+z}{10}\biggl)^{-1.2}.
\end{equation}
Eq. \ref{host_mass}-\ref{seed_mass} provide a very simple seeding prescription: at any redshift, both the mass of the halo and the mass of the BH that must be seeded in it are defined.

\subsection{Fraction of DM into PBHs}\label{fPBH}

In this section, we compute the minimum fraction of DM into PBHs ($f_{\rm PBH}$) required for our seeding mechanism to occur. The total DM density $\rho_{\rm DM}^{\rm tot}$ can be written, at any redshift, as the sum of two contributions: 
\begin{equation}\label{totDM}
\rho_{\rm DM}^{\rm tot}=\rho_{\rm DM}^{\rm h}+ \rho_{\rm DM}^{\rm IGM},
\end{equation}
where $\rho_{\rm DM}^{\rm h}$ and $\rho_{\rm DM}^{\rm IGM}$ is the DM density in the halos and in the IGM, respectively \citep[see also Sec. 2.1 in][]{Zip22}. In Sec. \ref{Sec:02}, we have argued that, in presence of small primordial non-Gaussianities, PBHs represent the {\it whole} of DM, but {\it only} in collapsed haloes. In other words, eq. \ref{totDM} can be re-written as:
\begin{equation}\label{totDM}
\rho_{\rm DM}^{\rm tot}=\rho_{\rm PBH}+ \rho_{\rm DM}^{\rm IGM},
\end{equation}
which implies
\begin{equation}\label{totDM}
f_{\rm PBH}=\frac{\rho_{\rm PBH}}{\rho_{\rm DM}^{\rm tot}}=\frac{\rho_{\rm DM}^{\rm tot}-\rho_{\rm DM}^{\rm IGM}}{\rho_{\rm DM}^{\rm tot}}=f_{\rm coll},
\end{equation}
where $f_{\rm coll}$ is the fraction of DM collapsed into halos. This quantity is defined in the Press-Schechter formalism according to the following expression: 
\begin{equation}
    f_{\rm coll}(z,\ M_h\geq M_{\rm min})=\mathrm{erfc}\biggl[\frac{\delta_c(z)}{\sqrt{2}\sigma_M(M_{\rm min})}\biggl],
\end{equation}
where $\delta_c(z)$ is the critical overdensity, $\sigma_M$ is the matter power spectrum, and the ratio $\nu=\delta_c/\sigma_M$ indicates the number of standard deviations to which an halo of mass $M_{\rm min}$ corresponds at a given redshift. 
We first compute to which $\sigma$-fluctuation corresponds the most stringent upper limit on $f_{\rm PBH}$ for $M_{\rm PBH}=30\ \rm M_{\odot}$, namely $f_{\rm PBH}\lesssim 10^{-4}$ \citep{Hutsi22}. We find that PBHs can constitute the {\it whole} DM into halos that corresponds to $\sigma$ fluctuations $\nu \gtrsim 3.7$ without violating current constraints of $f_{\rm PBH} (M_{\rm PBH}=30\ \rm M_{\odot}$).


Then, we quantify to which $\sigma$ fluctuation correspond the halos considered in Sec. \ref{seeds_form} at their crossing redshifts. In particular, we notice that the less extreme case is given by $M_{h,\times}=1.5\times 10^7~\rm M_{\odot}$ at $z_{\times}=22$. This combination corresponds to $\nu=4.5$ and $f_{\rm PBH}=3\times 10^{-6}$, the latter being our estimate for the minimum fraction of DM into PBHs required for our seeding mechanism to occur.

We finally note that, in the case of a uniform PBH spatial distribution (namely in the case of a perfectly Gaussian primordial power spectrum), the observationally allowed abundance of PBHs ($f_{\rm PBH}\leq 10^{-4}$) would prevent our seeding mechanism to happen due to the extremely low number of PBHs in the central region of DM halos.



\subsection{Implications for early SMBHs}\label{implications}
We use our model to interpret observations of early ($6<z<10$) SMBHs (Table \ref{table_obs}). We adopt estimates found in literature for the mass of the SMBH hosting halo. We then track $M_h$ backward in time, following \citep{Fakouri10}. Once that the crossing redshift ($z_{\times}$) and mass ($M_{h,\times}$) are determined, our model allows us to compute: the redshift ($z_{\rm seed}$), the mass ($M_{\rm seed}$), and the halo mass ($M_{ h,\rm seed}$) of the seed, as well as the average Eddington ratio $\langle \lambda_{\rm E}\rangle$ to which the seed should accrete to explain the observed SMBH masses (or X-ray luminosity, see below). 

In Fig. \ref{SEEDS-final}, we show both observational data and the seed masses resulting from our model (coloured as in Fig. \ref{Seedshalo_evo}). We find that $z\sim 6$ quasars can be explained with $6\times 10^4 M_{\rm \odot}$ seeds, planted at $z\sim32$, and growing at a sub-Eddington pace $\langle\lambda_{\rm E}\rangle\sim 0.55$. A similar scenario ($\langle\lambda_{\rm E}\rangle\sim 0.48$) can also reproduce the BH mass of GNz11 at $z=10.6$. However, we cannot reproduce the UV luminosity of this source, which is instead consistent with $\lambda_{\rm E}^{\rm obs}\sim 5.5$. We emphasize here that $\langle\lambda_{\rm E}\rangle$ must not be considered a proxy of $\lambda_{\rm E}^{\rm obs}$. In fact, while $\langle\lambda_{\rm E}\rangle$ provides an average value on the lifetime of the seed, $\lambda_{\rm E}^{\rm obs}$ represents the accretion rate of the BH at the time of the observation. Our model cannot predict the amplitude of variations around the average value, that are typically associated to the SMBH accretion process. Observations of $z\sim 6$ quasars \citep{farina2022, mazzucchelli2023} are consistent with variations of the order of $\Delta \lambda_{\rm E}/\langle\lambda_{\rm E}\rangle\sim 70$\% ($\langle\lambda_{\rm E}^{\rm obs}\rangle\sim 0.46\pm 0.32$). Whether variations of the accretion rate are occurring with similar amplitudes in earlier phases of BH growth is unknown. However, if this is the case, our estimate $\langle\lambda_{\rm E}\rangle\sim 0.48\pm 0.34$ implies that we can easily accomodate moderate episodes of super-Eddington accretion $\lambda_{\rm E}^{\rm obs}\sim 1-2$, while $\lambda_{\rm E}^{\rm obs}\sim 5.5$ values are difficult to reproduce (see also \citep{Bhatt}).

For what concerns UHZ1, since the BH mass is not known, we prefer to quantify the minimum $\langle \lambda_{\rm E}\rangle$ value that can reproduce the X-ray luminosity $L_X\geq2\times 10^{44}\ \rm ergs^{-1}$ estimated by \citep{Bogdan23}. UHZ1 data favour a scenario perfectly consistent with the one drawn by \citep{Bogdan23}: seeds must have been planted slightly later ($z\lesssim 22$) than the other two cases, are more massive ($1\times 10^5 M_{\rm \odot}$) and more efficiently accreting ($\langle\lambda_{\rm E}\rangle=0.9$), ending up into a $M_{\bullet}\sim4\times 10^7M_{\odot}$ at $z=10.1$. 

We use a similar approach for UHZ9. Due to the absence of a direct BH mass measurement, we quantify the minimum $\langle \lambda_{\rm E}\rangle$ value that can reproduce the bolometric luminosity $L_{\rm B}\gtrsim 10^{46}\ \rm ergs^{-1}$ estimated by \citep{GHZ9}. Similarly to UHZ1 case, slightly late ($z\sim 25$), and more efficiently accreting ($\lambda_E=0.96$) seeds are preferred. Starting from initial masses of $0.9\times 10^5\ M_{\odot}$, these seeds evolve into very massive BHs ($M_{\bullet}=1\times 10^{8}\ M_{\odot}$) at redshift $z=10$.  

By comparing the PBH seeding mechanism with other scenarios, we find that, since PBH seeds form at high redshift ($z\sim 20-40$) with high masses $M_{\rm seed}\sim 10^5\ M_{\odot}$, the mean accretion rate required to produce SMBHs at $z\sim 6-10$ is less extreme than the ones requested by the NSC (red box in Fig. \ref{SEEDS-final}) and PopIII (green box) scenarios, and comparable to the DCBH one (blue box).

\begin{center}
\begin{table}

\caption{Observations and predictions for $6<z<10$ SMBHs. For $z\sim 6$ quasars we consider $M_h=10^{13}M_{\odot}$ \citep{Costa24}. For UHZ1 (GNz11; UHZ9), given the estimated stellar mass of $M_*\simeq2.0\times 10^8 M_{\odot}$ ($M_*\simeq 6\times 10^8 M_{\odot}$; $M_*\simeq 3.3 \times 10^{8} M_{\odot}$ \citep{Atek23, Castellano22}), and assuming $M_*=\epsilon_* (\Omega_b/\Omega_m) M_h$, with a star formation efficiency $\epsilon_*=0.1$, we find $M_h=2.0\times 10^{10} M_{\odot}$ ($M_h=2\times 10^{11}M_{\odot}$; $3.3\times 10^{10} M_{\odot}$). For $\lambda_{\rm E^{obs}}$ we report the values from \citep{farina2022,  mazzucchelli2023} and \citep{GNZ11}(GNz11).} 
\centering

\setlength{\tabcolsep}{3.8pt}
\begin{tabular}{l c c c c}

 \hline
  &Quasars&UHZ1&GNz11&UHZ9 \\ 
  \hline
  \vspace{0.05mm}\\
  \multicolumn{2}{l}{OBSERVATIONS}\\  
  \vspace{0.05mm}\\

 \hline
 $z$& 6 & 10.1&10.6&10\\
 $M_{\bullet}[M_\odot]$ & $10^{9}$ & 4$\times 10^7$&2$\times 10^6$&$10^8$\\
 $M_h[M_\odot]$ & $10^{13}$ & 2.0$\times 10^{10}$ & $2\times 10^{11}$& $2\times 10^8$\\
 $\lambda_{\rm E}^{\rm obs}$ &$0.46\pm 0.32$&-&5.5&-\\
 \hline
 \vspace{0.05mm}\\
 \multicolumn{2}{l}{PREDICTIONS}\\ 
 \vspace{0.05mm}\\
 \hline
$z_{\rm seed}$&  $32$& $21.7$&$27.5$& $25$\\
$M_{\rm seed}[M_\odot]$&  $6\times10^4 $&$1.1\times10^5 $& $0.8\times10^5 $& $0.9\times10^5 $\\
$M_{h, \rm  seed}[M_\odot]$ &$4\times10^7\ $&$1.5\times10^8\ $&$0.7\times10^8$&$0.7\times10^8$\\
$\langle \lambda_{\rm E}\rangle$ &0.55&0.92&0.48&0.96\\

 \hline

\end{tabular}
\label{table_obs}

\end{table}
\end{center}

\section{Summary and discussion}\label{Sec:04}
We have considered a cosmological framework in which the primordial power spectrum is characterised by local non-Gaussianities that are small enough not to violate CMB constraints. In this context, primordial black holes (PBHs) are initially clustered and preferentially formed in the high-$\sigma$ fluctuations of the large-scale density field, out of which dark matter (DM) halos are originated.

We have shown that runaway mergers of  PBHs in the central region ($r<1$~pc) of  high redshift ($20<z<40$) halos can lead to the formation of  massive black hole seeds. This mechanism provides a new route to rapid SMBH growth.   The main results of this study are:

\begin{figure*}
\includegraphics[width=1\textwidth]{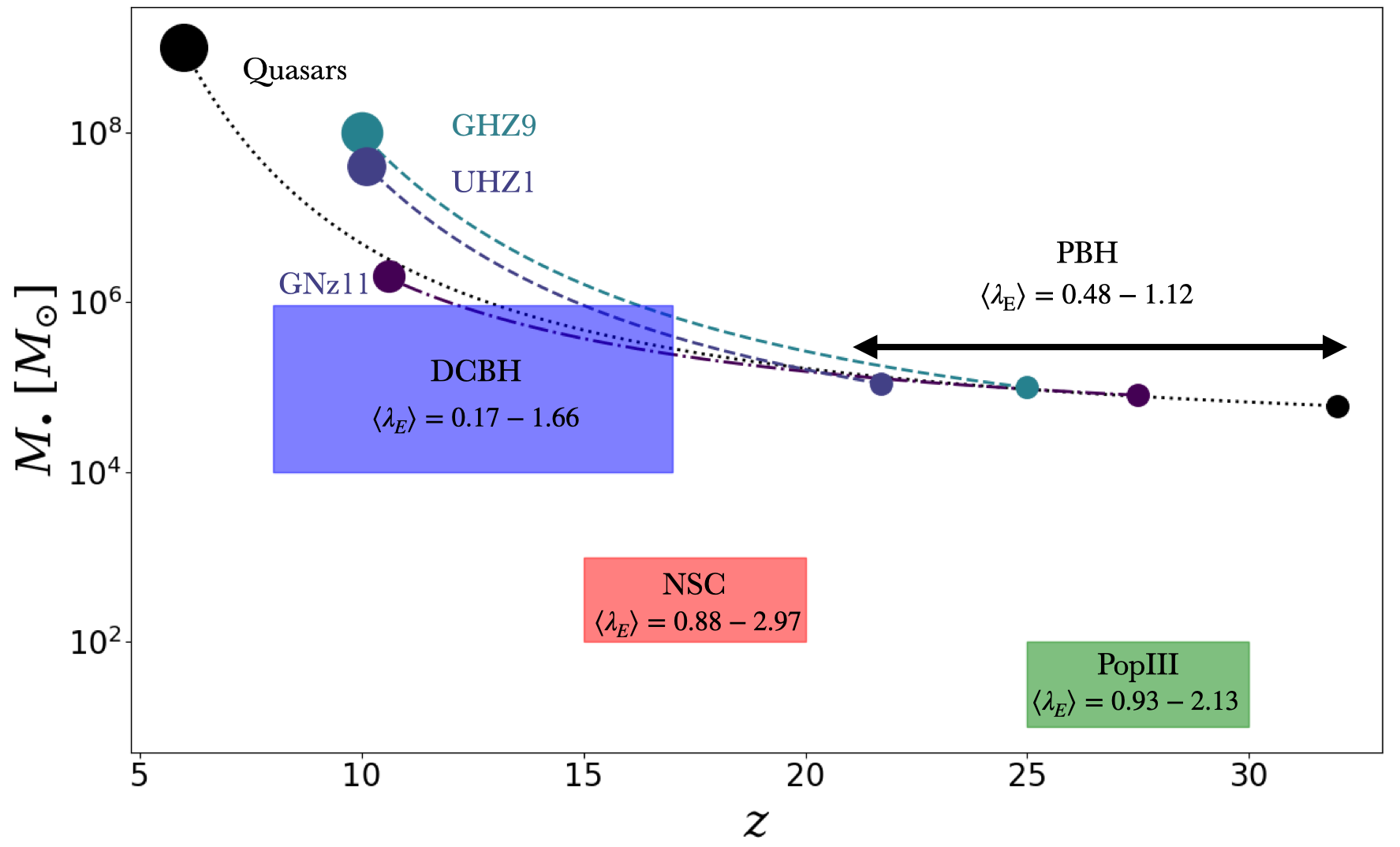}
\caption{Growth of PBH seeds accreting into $z\sim 6-10$ SMBHs. Large (small) circles represent SMBH observations (predicted PBH seeds). By assigning the halo mass to the observed SMBH hosts and evolving it back in cosmic time, we link each SMBH to its PBH progenitor. Dashed and dotted lines represent the assembly of the SMBHs assuming the constant mean accretion rate $\langle \lambda_E \rangle$ reported in Tab. \ref{table_obs}. We find that PBH seeds can explain the masses of $z\sim 6-10$ SMBHs with $0.48<\langle \lambda_E \rangle<1.12$ (see Fig. \ref{SEEDS-triple} for further details). Colored boxes report the redshift and mass ranges for different seeding scenarios: $z=8-17$ and $M_{\rm seed}=10^{4-6}\ M_{\odot}$ for DCBHs \citep[blue box,][]{Ferrara14}; $z=15-20$ and $M_{\rm seed}=10^{2-3}\ M_{\odot}$ for NSCs \citep[red box,][]{Volonteri03}; $z=25-30$ and $M_{\rm seed}=10^{1-2}\ M_{\odot}$ for PopIII remnants  \citep[green box][]{Madau01}. For each seeding scenario, we report the minimum and maximum $\langle \lambda_E \rangle$ value required to explain the observed masses of $z\sim 6-10$ SMBHs. The PBH seeding mechanism combines the early birth ($z\gtrsim20$) of \textit{light}/\textit{intermediate} seeds with the high masses ($M_{\rm seed}\sim 10^5\ M_{\odot}$) of \textit{heavy} seeds. Thus, PBH seeds can explain $z\sim 6-10$ SMBH observations by growing with less extreme $\langle \lambda_E \rangle$ values. 
}\label{SEEDS-final}
\end{figure*}

\begin{itemize}

\item Due to dynamical friction on dense gas in early dark matter halos, PBHs  of mass $\simeq 30\, M_\odot$ tend to concentrate into a compact ($ r_c\sim10^{-3}\ \rm pc$) core containing  $\simeq 1000$ PBHs. At such high concentration,  PBHs form binaries triggering their runaway merger process, eventually leading to a massive ($10^{4-5}~M_{\rm \odot}$) BH seed. 

\item Massive BH seeds predominantly form in early  ($6<z<30$), rare halos of mass $5\times 10^7-10^9~M_{\rm \odot}$, representing  $\gtrsim 5\sigma$ fluctuations of the density field. Based on these results we derive a physical seeding prescription (eq. \ref{host_mass} and eq.  \ref{seed_mass}) that can be used in theoretical and numerical studies. 
\item The seed masses predicted by our model are large enough to explain recent JWST observations of early SMBHs  \citep{GNZ11, Bogdan23, GHZ9} without the need for super-Eddington accretion. Moreover, our predictions nicely agree with the observed properties of  $z\sim6$ quasars, matching at the same time the host dark matter halo ($M_h\sim10^{12-13}\ M_{\odot}$), and SMBH ($M_{\bullet}\sim 10^{8-10}\ M_{\odot}$) masses with a conservative mean accretion rate, $\langle \lambda_E\rangle \sim 0.5$.
\item With a minimal fraction of PBHs in DM ($f_{\rm PBH}\sim 3 \times 10^{-6}$), our seeding mechanism combines the early birth ($z\gtrsim20$) of \textit{light}/\textit{intermediate} seeds with the high masses ($M_{\rm seed}\sim 10^5\ M_{\odot}$) of \textit{heavy} seeds, resulting in less stringent requirements on the BH accretion history.
\item Our model predicts that thousands of PBH-PBH mergers occur in the runaway phase, thus resulting into the emission of copious gravitational waves (GWs). We defer to a future work a calculation of the resulting GW signal, its contribution to the stochastic GW background, and detectability with future GW instruments, such as the Einstein Telescope \citep{ET}. If the GW signal resulting from the proposed seed formation process comes out to be detectable, it will be suitable to constrain primordial non-gaussianities and therefore to constrain inflationary models.
\end{itemize}

Although we consider our results as fairly robust, a number of assumptions made deserve further scrutiny. 
In Sec. \ref{Sec:02}, we have assumed that star formation is suppressed in the central regions of the halo. This assumption is based on two heuristic arguments:  (i) competitive accretion by PBHs should rapidly devoid the gas, thus strongly inhibiting its conversion into stars; (ii) UV radiation produced by PBH gas accretion should largely photo-dissociate $H_2$ molecules, again preventing star formation. 

It is worth noting, though, that the proposed seeding mechanism might work equally well even in the presence of stars. Although we considered only dynamical friction on the gas, the inclusion of a stellar component would also contribute to friction. For a standard 1-100 $M_\odot$ Salpeter stellar IMF \citep{Salpeter55}, $ 30\ M_{\odot}$ PBHs are heavier than $\gtrsim 99$\% of the stars. We then expect PBHs to sink towards the halo center, similarly to the gas-only case, by kicking out lighter stars.

We have also neglected the effect that minor and major mergers of DM halos could have on the PBH core formation process. DM halo mergers modify the halo density profile, and may perturb the orbits of PBHs as they sink towards the halo center. As for minor mergers, it has been shown that their effect is to smoothly feed the halo outskirts without significantly affecting the halo central regions \citep{Salvador98}. Major mergers might be in principle more disruptive, as they can alter even the central distribution of dark matter.  If so, they could possibly hamper the settling of the PBH core.  However,  the physical process proposed here is effective only in high-$\sigma$ density fluctuations, for which major mergers are rare. In other words, major mergers are likely affecting the number of seeds produced by our model, but they should not prevent their formation entirely.  Nevertheless, a quantitative assessment of major merger effects on the results presented in this work will require dedicated numerical simulations.

\section*{Data Availability}
Data generated in this research will be shared on reasonable request to the corresponding author.

\section*{ACKNOWLEDGEMENTS}
FZ thanks A. Mesinger for for numerous discussions that contributed to understanding the framework essential for developing this work; A. Pallottini for the useful criticism that led to improvements of the model; M. Volonteri for a useful discussion that deepened the understanding of some of the processes described in this work. FZ also thanks G. Franciolini, M. Pieroni, S. Matarrese, A. Ricciardone, A. Riotto for useful discussions.
SG acknowledges support from the PRIN 2022 project (2022TKPB2P), titled: "BIG-z: Building the Giants: accretion, feedback and assembly in $z>6$ quasars". AF acknowledges support from the ERC Advanced Grant INTERSTELLAR H2020/740120. Partial support from the Carl Friedrich von Siemens-Forschungspreis der Alexander von Humboldt-Stiftung Research Award is kindly acknowledged. We gratefully acknowledge computational resources of the Center for High Performance Computing (CHPC) at SNS.


\providecommand{\href}[2]{#2}\begingroup\raggedright\endgroup

\appendix
\section{Comparison with RAPSTER}\label{Rapster}
According to our model, the seed mass $M_{\rm seed}$ is equal to the core mass $M_c$ when Eq. \ref{critical} is satisfied. We remind that this condition means that all the PBHs contained in the core can collapse into a single seed if they are retained within the core itself during the runaway mergers. This occurs when the mass and the radius of the core are such that its escape velocity is larger than the maximum kick velocity $v_k^{max}$ that a PBH can receive during the runaway mergers. 
This value can be as small as $\simeq 180\ \rm km\ s^{-1}$, in the case of non-spinning PBHs \citep{Gonzalez07}, and as large as $\simeq 4000\ \rm km\ s^{-1}$ \citep{Baker08} for an optimal mass ratio and spin configuration. Since our assumed $v_k^{max}=1000\ \rm km\ s^{-1}$ value is smaller than the latter, we may underestimate the number of PBHs that are lost because of GW recoils. Furthermore, we are neglecting the fact that in a binary merger the remnant mass is lower then the sum of the two black holes masses, since a fraction of the initial rest mass is converted into GW emission. In other words, by considering $M_{\rm seed}=M_c$, we may overestimate the final seed mass.

To proper compute the fraction of PBHs retained by the cluster during the runaway mergers  and the mass loss due to GW emissions, we adopt and modify the publicly available code Rapid Cluster Evolution \citep[RAPSTER,][]{Rapster}. RAPSTER follows (i) the BH formation process from the death of massive stars, (ii) the formation of a BH core in the cluster, (iii) the subsequent dynamical formation of binary black holes (BBHs), and (iv) the final merging of BBHs in the cores of nuclear star clusters; it finally provides (v) the properties of the system resulting from the BBH merging and computes the corresponding gravitational wave emission. In particular, RAPSTER accounts for the mass loss due to recoil kicks, three body interactions, and GW emission and thus allows us to properly compute the mass retained in the cluster during the collapse (hereafter $M_{rap}$). The aim of this Appendix is thus to compare our estimated seed mass $M_{\rm seed}$ with the actual value $M_{rap}$.

The RAPSTER calculations start from an initial configuration that consists of a giant molecular cloud (GMC) of mass $M_{cl,0}$, radius $r_{cl,0}$ and metallicity $Z$. A fraction of the GMC mass, that depends on the assumed star formation efficiency ($\varepsilon_{\star}=0.1$), fragments into stars whose masses are distributed according to the assumed initial mass function\footnote{The Kroupa IMF \citep{Kroupa01} is adopted, with a $-2.3$ power law index for $M_{\star}>1\ M_{\odot}$.} (IMF). Massive stars ($>20~M_{\odot}$) evolve into BH remnants. This process determines the initial number of BHs in the cluster, $N_{\rm BH}^{tot}$. These black holes proceed to form a denser core inside the cluster due to energy equipartition between stars and black holes in a process called mass segregation. The radius of this core is computed as:
\begin{equation}\label{r_core}
    \frac{r_{c, bh}}{r_{cl,0}}=0.02\frac{8}{\xi_{min}}\biggl( \frac{M_{\bullet}}{10\  M_{\odot}}\biggl)^2\frac{n_{\rm bh}}{1000}\frac{0.64 M_{\odot}}{\bar{m}}\frac{10^6 M_{\odot}}{M_{cl}},
\end{equation}
where $\xi=M_{\bullet}\sigma^2_{\rm bh}/ \bar{m}\sigma^2_{\star}$, is the temperature ratio between black holes and stars, $\sigma_{\star}$ is the stellar velocity dispersion, and $\bar{m}$ is the mean stellar mass.  As a result of interactions that occur in the dense
cluster, energy is distributed among the members of the
system, with BHs slowing down and lighter objects gaining kinetic energy via two-body encounters. As relics of massive star evolution, BHs become the heaviest components in the cluster, and if not already formed in the core, they sink into the central regions via dynamical friction on the stars. 

We adapt the RAPSTER code to follow the evolution of a core composed of PBHs that is formed due to the dynamical friction on the gas. To this aim, we use the radius ($r_c$) and mass ($M_c=N_{\rm PBH}M_{\rm PBH}$) of the core (as computed in Sec. \ref{Sec:02.5}) for the initial configuration ($r_{cl,0}$ and $M_{cl,0}$ in the RAPSTER formalism). 

We fix the cluster position in the center of the galaxy and we remove stars from the computation since we expect feedback from PBH accretion to stop the star formation process. 
We finally initialize PBHs spin to a monochromatic distribution peaked at the 0 value, as we expect PBHs to be non-spinning before the first generation of mergers \citep{DeLuca20}.

\begin{figure}
\begin{center}
\includegraphics[width=0.8\textwidth]
{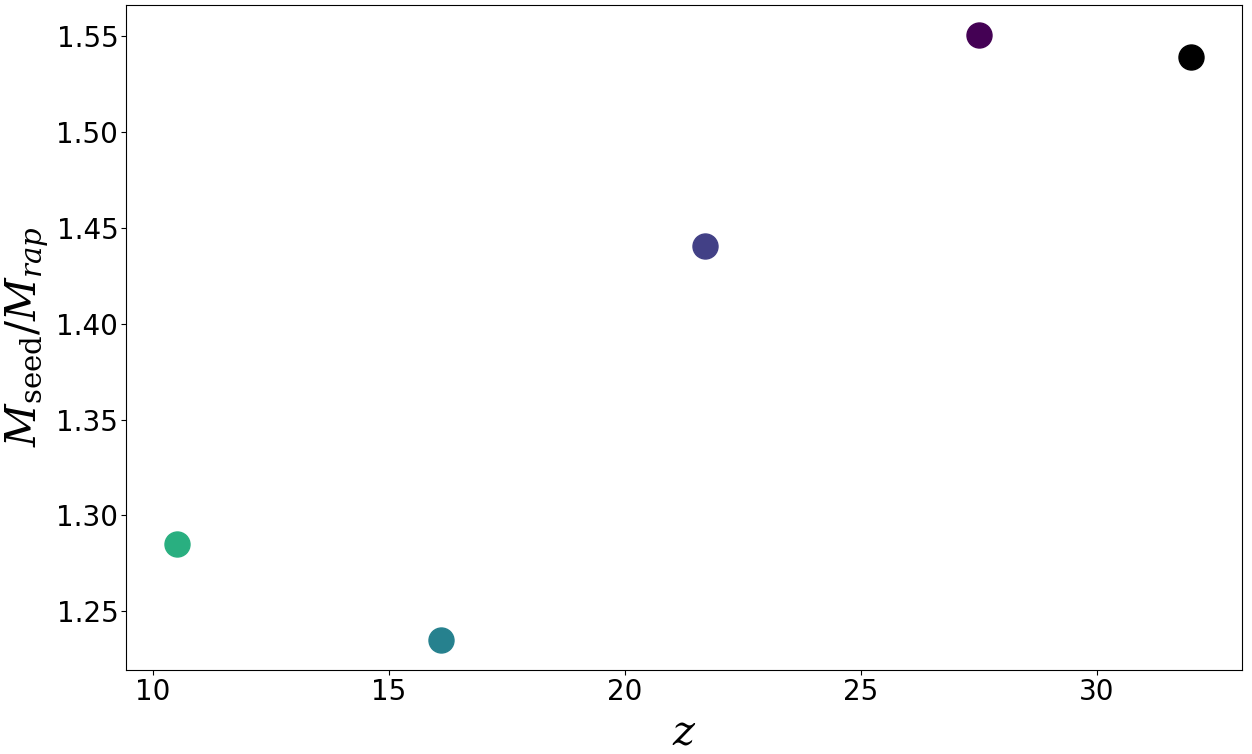}
\caption{Seed mass comparison. Ratio between our ($M_{\rm seed}$) and RAPSTER ($M_{rap}$) estimate of the final seed mass as a function of redshift, for the same cases shown in Fig. \ref{Seedshalo_evo}.
}\label{Masse_comp}
\end{center}
\end{figure}

We plot in Fig. \ref{Masse_comp} the $M_{\rm seed}$ to $M_{rap}$ ratio for seeds formed at different redshifts. We find that, as expected, $M_{\rm seed}$ is always overestimating the $M_{rap}$ value; however, the difference between the two values is not significant, being always smaller that a factor 1.6. We thus conclude that the equation $M_{\rm seed}=M_c$ and our assumption of $v_k^{max}=1000$~$\rm km\ s^{-1}$ provide a satisfactory estimate of the final seed mass.

\section{Observational uncertainties}\label{AppB}

\begin{figure*}
\includegraphics[width=1\textwidth]{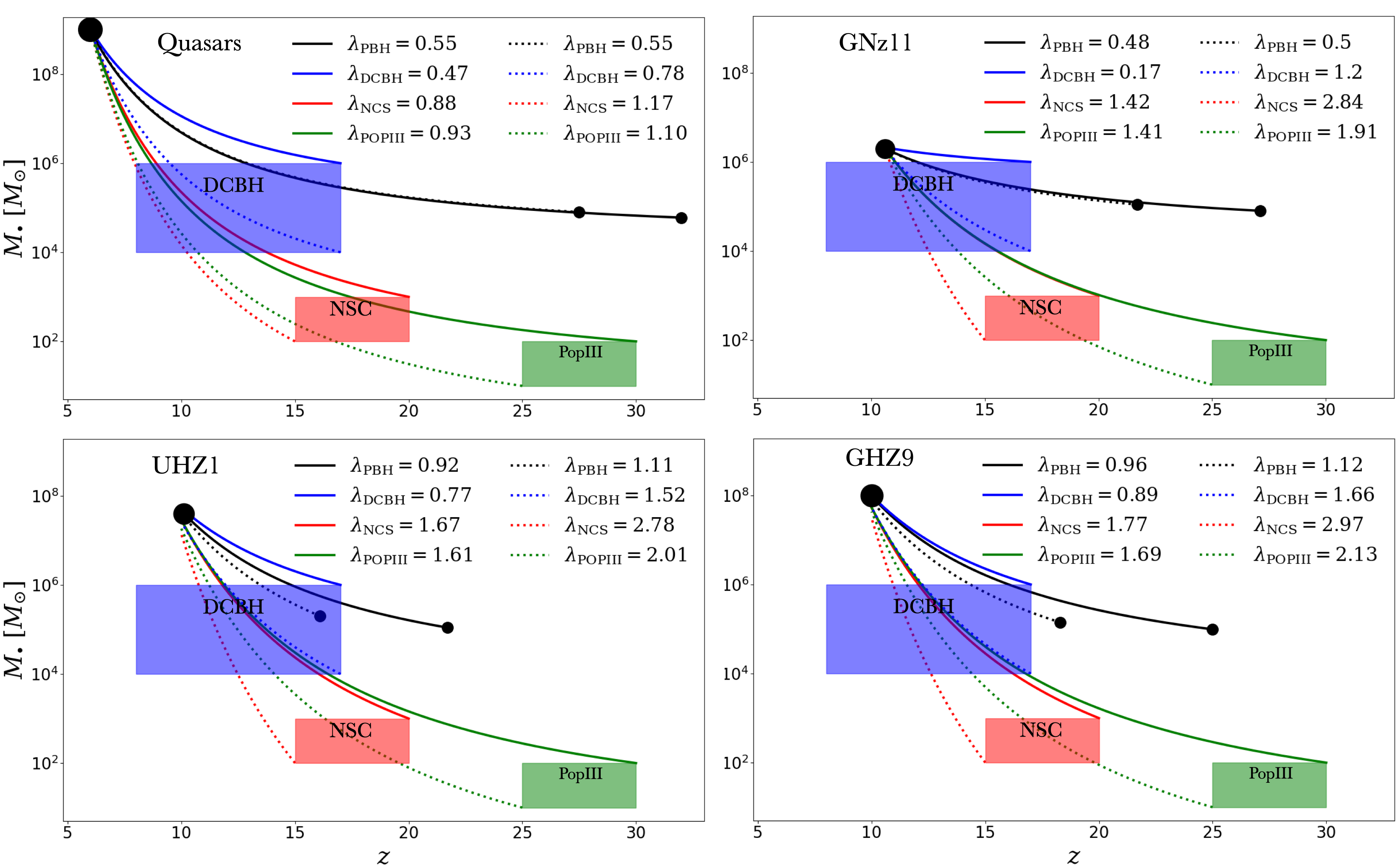}
\caption{Redshift evolution of seeds into $z\sim 6-10$ SMBHs. Large (small) circles represent SMBH observations (predicted PBH seeds). By assigning the halo mass to the observed SMBH hosts and evolving it back in cosmic time, we link each SMBH to its progenitor. The continuous (dotted) line represents the minimum (maximum) mean accretion rate ($  \langle \lambda_E \rangle$) necessary to reach the final SMBH mass. The colored boxes mark the masses and redshifts interval of different seeding scenarios: DCBH (blue box), NSC (red box), PopIII remnants (green box).
\textit{Upper left panel:}
The black circle represent the observations of a typical SMBH ($ M_{\bullet}=10^{9}\ M_{\odot}$) at redshift $z=6$. PBH Seed masses of $M_{\rm seed}=6-8\times 10^4 M_{\odot}$ planted at $z=27-32$ can reproduce the observed masses if accreting at a mean sub-Eddington accretion rate ($\lambda_E=0.55$). For other seeding scenarios we find: $\lambda_{\rm DCBH}=0.47-0.78$, $\lambda_{\rm NSC}=0.88-1.17$, $\lambda_{\rm PopIII}=0.93-1.10$. \textit{Upper right panel:} The large black dot represents the massive black hole ($ M_{\bullet}= 2\times 10^{6}\ M_{\odot}$, \citep{GNZ11}) hosted by GNz11 at redshift $ z=10.6$. We find that a seed mass of $M_{\rm seed}=0.8-1\times 10^5\ M_{\odot}$ planted at $z=21.7-27.5$ can reproduce the observed SMBH mass ($M_{\bullet}\sim 2\times 10^6\ M_{\odot}$) with a mean sub-Eddington accretion rate ($\lambda_E=0.48-0.50$). For other seeding mechanism we find: $\lambda_{\rm DCBH}=0.17-1.20$, $\lambda_{\rm NSC}=1.42-2.84$, $\lambda_{\rm PopIII}=1.41-1.91$. 
\textit{Lower left panel:}
The large black dot represents the massive black hole ($ L_X\sim 2\times 10^{44}\ \rm erg\, s^{-1}$, \citep{UHZ1}) powering UHZ1 at redshift $ z=10.1$. PBHs seeds of mass $M_{\rm seed}\sim 10^5~\rm M_{\odot}$ planted at $z=16.1-21.7$ can reproduce the observed luminosity with a mean accretion rate of $\lambda_E=0.92-1.11$, and reach a final mass of $3-4\times 10^7\ M_{\odot}$. Applying the same methodology to other seeding scenarios we find: $\lambda_{\rm DCBH}=0.77-1.52$, $\lambda_{\rm NSC}=1.67-2.78$, $\lambda_{\rm PopIII}=1.61-2.01$. \textit{Lower right panel:} The large black dot represents the massive black hole hosted in GHZ9 at redshift $z=10$. PBH seeds of mass $M_{\rm seed}\sim 10^5~\rm M_{\odot}$ planted at redshift $z=18.3-25$ in massive halos ($M_h=0.7-2\times 10^8\ M_{\odot}$), and accreting with $\lambda_E=0.96-1.12$ reproduce the observed bolometric luminosity ($L_{\rm B}\simeq 10^{46}\ \rm erg\,s^{-1}$, \citep{GHZ9}) and stellar mass $M_{\star}=0.5-3.3\times 10^{8}\ M_{\odot}$ \citep{Atek23, Castellano22}, and reach final BH masses of $M_{\bullet}=0.7-1\times 10^8\ M_{\odot}$. Similarly, for other seeding scenarios we find: $\lambda_{\rm DCBH}=0.89-1.66$, $\lambda_{\rm NSC}=1.77-2.97$, $\lambda_{\rm PopIII}=1.69-2.13$.
}\label{SEEDS-triple}
\end{figure*}

In this Appendix, we evaluate the impact of observational uncertainties on the mean accretion rates predicted for the PBH seeds in the four cases discussed in Sec. \ref{Sec:04}. 

In Fig. \ref{SEEDS-final}, we have connected each observational data (large circles) to a single PBH seed (small circles) with a single mean accretion rate $\langle\lambda_{\rm E}\rangle$. We remind that this has been done through the following steps: i) to estimate the mass of the hosting halo ($M_{h}$) from the stellar mass ($M_{\star}$) inferred from observations: $M_{\star}=\varepsilon_{\star}(\Omega_b/\Omega_m) M_{h}$, where $\varepsilon_{\star}=0.1$ is the star formation efficiency; ii) to track $M_{h}$ backward in time, following \citep{Fakouri10}; iii) to identify the crossing redshift ($z_{\times}$) and mass ($M_{h,\times}$) that determine the redshift ($z_{\rm seed}$) and mass ($M_{\rm seed}$) of the PBH seed; iv) to compute the average Eddington ratio to which the seed should accrete to explain the observed data. 
So far, for step i), we have used the $M_{\star}$ upper limit of each observational case. 

In this Appendix, we instead consider both the lower and upper limits of $M_{\star}$ that convert into a minimum and maximum $z_{\rm seed}$ and $M_{\rm seed}$, thus finally providing a maximum and minimum value for $\langle\lambda_{\rm E}\rangle$, respectively.
Similarly, for the other seeding scenarios, we consider the minimum/maximum seed mass and formation redshift as predicted by theoretical models. In these cases, the minimum and maximum $\langle\lambda_{\rm E}\rangle$ value is associated to the maxima seed mass/formation redshift and minima seed mass/formation redshift, respectively. The results are shown in Fig. \ref{SEEDS-triple}.

For what concerns $z\sim 6$ quasars (upper left panel), we find that their masses can be explained without relying on super-Eddington accretion, independently of the considered seeding mechanism. Following \citep{Costa24}, we assume that $M_{\bullet}=10^9\ M_{\odot}$ are hosted into $M_h=10^{12-13}\ M_{\odot}$ dark matter halos. These halo masses correspond to PBH progenitors with $M_{\rm seed}=6-8\times 10^4 M_{\odot}$ planted at redshift $z=32-27$. To grow into the SMBHs at redshift $z=6$, both seeds require the same accretion history with a mean accretion rate of $\lambda_{\rm PBH}=0.55$. This value is consistent with the accretion rate required for DCBHs to evolve into the same SMBH masses $\lambda_{\rm DCBH}=0.47-0.78$. Viceversa, both the NSCs and PopIII scenarios require accretion rates closer to the Eddington limit: ($\lambda_{\rm NSC}=0.88-1.17$, $\lambda_{\rm PopIII}=0.93-1.10$).

To determine the mass of the dark matter halo hosting GNz11 (upper right panel), we use the observational esteem of the galaxy stellar mass $M_{\star}\simeq10^{9-10}\ M_{\odot}$ by \citep{GNZ11}. Thus, PBH seeds forming between redshift $z=27.5-21.7$ with masses of $M_{\rm seed}=0.8-1\times 10^5\ M_{\odot}$ reproduce the observed mass of the BH inside GNz11 ($M_{\bullet}=2\times10^6\ M_{\odot}$) with a mean accretion rate of $\lambda_{\rm PBH}\sim 0.48-0.50$. For what concerns the DCBH scenario, models predict seed masses very close to the BH mass in GNz11, and an interval of redshift for their formation that includes the GNz11 redshift. In this case, we only consider the upper limit of this interval that implies an extremely low mean accretion rate value ($\lambda_{\rm DCBH}=0.17$). In the other cases, super-Eddington accretion is always necessary to match the observed BH mass: $\lambda_{\rm NSC}=1.42-2.84$, $\lambda_{\rm PopIII}=1.41-1.91$.

In the case of UHZ1 and GHZ9 \citep{UHZ1, GHZ9} the inferred stellar masses are very close to the BH masses. These high BH to stellar mass ratios ($M_{\bullet}/M_{\star}\sim 1$) imply a very rapid BH assembly for both sources. For UHZ1 (lower left pabel), we adopt the stellar mass estimate ($M_{\star}=0.4-1.9\times 10^8\ M_{\odot}$) provided by \citep{Castellano22} which converts into $M_h\simeq0.4-2\times 10^9\ M_{\odot}$. PBH seeds with masses $M_{\rm seed}\simeq 10^5 M_{\odot}$ planted at redshift $z=16.1-21.7$ can reproduce the observed luminosity ($L_x\sim2\times10^{44} \rm{erg\ s^{-1}}$) with a mean accretion rate of $\lambda_{\rm PBH}=0.92-1.11$. Similarly, DCBHs can reproduce the UHZ1 luminosity with $\lambda_{\rm DCBH}=0.77-1.52$. The other two seeding scenarios require instead higher (super-Eddington) mean accretion rates for a prolonged period of time ($<200\ \rm Myr$): $\lambda_{\rm NSC}=1.67-2.78$, $\lambda_{\rm PopIII}=1.61-2.01$.

In the case of GHZ9 (lower right panel), for the galaxy stellar mass we use the value $M_{\star}=0.5-3.3\times 10^8\ M_{\odot}$ \citep{Castellano22, Atek23}. PBH seeds planted at redshift $z=18-25$ with masses $M_{\rm seed}=10^5\ M_{\odot}$ can explain the observed luminosity ($L_B=10^{46} \rm{erg\ s^{-1}}$) with a mean accretion history of $\lambda_{\rm PBH}=0.96-1.12$. In the case of DCBHs, a mean accretion rate close to the Eddington rate ($\lambda_{\rm DCBH}=0.89-1.66$) is required to explain the high inferred BH mass ($M_{\bullet}\sim 10^8\ M_{\odot}$) at the high observational redshift ($z=10.3$). The other two seeding scenarios require sustained super-Eddington accretion: $\lambda_{\rm NSC}=1.77-2.97$, $\lambda_{\rm PopIII}=1.69-2.13$.

\label{lastpage}

\end{document}